\documentclass[aps,prb,twocolumn,superscriptaddress,showpacs]{revtex4}
\usepackage{graphicx}
\usepackage{longtable}
\usepackage{array}
\usepackage{braket}
\newcolumntype{L}{>{\centering\arraybackslash}m{2cm}}
\newcolumntype{R}{>{\centering\arraybackslash}m{1.5cm}}
\newcolumntype{K}{>{\centering\arraybackslash}m{1.3cm}}
\usepackage{gensymb} % or simply amstext
\newcommand{\angstrom}{\text{\normalfont\AA}}

\usepackage{xcolor}

\begin{document}

% Use the \preprint command to place your local institutional report
% number in the upper righthand corner of the title page in preprint mode.
% Multiple \preprint commands are allowed.
% Use the 'preprintnumbers' class option to override journal defaults
% to display numbers if necessary
%\preprint{}

%Title of paper
\title{Gapless quantum excitations from an Ice-like Splayed Ferromagnetic ground state in stoichiometric $\bf{Yb_2Ti_2O_7}$.}

\author{J. Gaudet}
%\email{gaudej@mcmaster.ca}
\affiliation{Department of Physics and Astronomy, McMaster University, Hamilton, ON L8S 4M1 Canada}

\author{K. A. Ross}
\altaffiliation[Current Address: ]
    {Colorado State University, Fort Collins, Colorado 80523-1875, U.S.A.}
\affiliation{Institute for Quantum Matter and Department of Physics and Astronomy, Johns Hopkins University, Baltimore, Maryland 21218, USA}
\affiliation{NIST Center for Neutron Research, National Institute of Standards and Technology, Gaithersburg, Maryland 20899, USA}

 \author{E. Kermarrec}
% %\email{edw.kermarrec@gmail.com}
\affiliation{Department of Physics and Astronomy, McMaster University, Hamilton, ON L8S 4M1 Canada}

\author{N. P. Butch}
%\email{rosska2@pha.jhu.edu}
\affiliation{NIST Center for Neutron Research, National Institute of Standards and Technology, Gaithersburg, Maryland 20899, USA}

\author{G. Ehlers}
%\email{rosska2@pha.jhu.edu}
\affiliation{Quantum Condensed Matter Division, Oak Ridge National Laboratory, Oak Ridge, Tennessee 37831, USA}

\author{H. A. Dabkowska}
%\email{rosska2@pha.jhu.edu}
\affiliation{Brockhouse Institute for Materials Research, Hamilton, ON L8S 4M1 Canada}

\author{B. D. Gaulin}
%\email{bruce.gaulin@gmail.com}
\affiliation{Department of Physics and Astronomy, McMaster University, Hamilton, ON L8S 4M1 Canada}
\affiliation{Brockhouse Institute for Materials Research, Hamilton, ON L8S 4M1 Canada}
\affiliation{Canadian Institute for Materials Research, 180 Dundas Street West, Toronto, Ontario M5G 1Z8, Canada}

% repeat the \author .. \affiliation  etc. as needed
% \email, \thanks, \homepage, \altaffiliation all apply to the current
% author. Explanatory text should go in the []'s, actual e-mail
% address or url should go in the {}'s for \email and \homepage.
% Please use the appropriate macro foreach each type of information

% \affiliation command applies to all authors since the last
% \affiliation command. The \affiliation command should follow the
% other information
% \affiliation can be followed by \email, \homepage, \thanks as well.

%\email[]{Your e-mail address}
%\homepage[]{Your web page}
%\thanks{}
%\altaffiliation{}

%Collaboration name if desired (requires use of superscriptaddress
%option in \documentclass). \noaffiliation is required (may also be
%used with the \author command).
%\collaboration can be followed by \email, \homepage, \thanks as well.
%\collaboration{}
%\noaffiliation

\date{\today}

\begin{abstract}
The ground state of the quantum spin ice candidate magnet Yb$_2$Ti$_2$O$_7$ is known to be 
sensitive to weak disorder at the $\sim$ 1 $\%$ level which occurs in single crystals grown from the melt.  Powders produced by solid state synthesis tend to be stoichiometric and display large and sharp heat capacity anomalies at relatively high temperatures, T$_C$$\sim$ 0.26 K.  We have carried out neutron elastic and inelastic measurements on well characterized and equilibrated stoichiometric powder samples of Yb$_2$Ti$_2$O$_7$ which show resolution-limited Bragg peaks to appear at low temperatures, but whose onset correlates with temperatures much higher than T$_C$.  The corresponding magnetic structure is best described as an ice-like splayed ferromagnet.   The spin dynamics in Yb$_2$Ti$_2$O$_7$ are shown to be gapless on an energy scale $<$ 0.09 meV at all temperatures, and organized into a continuum of scattering with vestiges of highly overdamped ferromagnetic spin waves present.  These excitations differ greatly from conventional spin waves predicted for Yb$_2$Ti$_2$O$_7$'s mean field ordered state, but appear robust to weak disorder as they are largely consistent with those displayed by non-stoichiometric crushed single crystals and single crystals, as well as by powder samples of Yb$_2$Ti$_2$O$_7$'s sister quantum magnet Yb$_2$Sn$_2$O$_7$.
\end{abstract}

%\keywords{neutron scattering, crystal-fields, rare earth titanates, stuffing}

% insert suggested PACS numbers in braces on next line 
% insert suggested keywords - APS authors don't need to do this
%\keywords{}

%\maketitle must follow title, authors, abstract, \pacs, and \keywords
\maketitle

% body of paper here - Use proper section commands
% References should be done using the \cite, \ref, and \label commands

%%%%%%%%%%%%%%%%%%%%%%%%%%%%%%%%%%%%%%%%%%%%%%%%%%%%%%%%%%%%%%%%%%%%%%%%%%%%%%%%%%%%%%

\section{INTRODUCTION}
Pyrochlore magnets of the form A$_2$B$_2$O$_7$ have been of great topical interest as both the A and B sublattices independently form networks of corner-sharing tetrahedra, one of the canonical crystalline architectures supporting geometrical frustration in three dimensions~\cite{Greedan,Subramanian}.  The cubic rare earth titanates, of the form RE$_2$Ti$_2$O$_7$ have been especially relevant as many magnetic RE$^{3+}$ ions can occupy the A-site of the structure and where the non-magnetic Ti$^{4+}$ occupy the B-site. This pyrochlore family can also be relatively easily produced in both powder and single crystal form~\cite{Balakrishnan1998,Dabkowska2010,Prabhakaran2011}.  One of the family members, Yb$_2$Ti$_2$O$_7$, has been particularly topical as it has been proposed as a realization of quantum spin ice~\cite{Chang2012,GingrasRev,Yasui2003,Cao2009,Gardner2004,Dalmas2006,RossPRX,Ross2011,Ross2009,Thompson2011,Applegate2012}.  It displays a net ferromagnetic  Curie-Weiss constant of $\sim$ 0.6 K~\cite{Bramwell2000,Hodges2001}, and crystal field (CF) effects give rise to a CF ground state doublet at the Yb$^{3+}$ site made up of primarily m$_J$=$\pm$ 1/2 eigenvectors and local XY anisotropy~\cite{Malkin2004,Bertin2012,Gaudet2015}.  Yb$_2$Ti$_2$O$_7$ is therefore a good realization of quantum S$_{\rm{eff}}=1/2$ spins decorating a network of corner-sharing tetrahedra - the pyrochlore lattice.

The microscopic spin Hamiltonian for Yb$_2$Ti$_2$O$_7$ has been estimated using neutron spectroscopic measurements of spin waves in its high field polarized state~\cite{RossPRX}.  While the zero field phase of Yb$_2$Ti$_2$O$_7$ does not show well defined spin waves in single crystals, a magnetic field applied along the [1-10] direction pushes Yb$_2$Ti$_2$O$_7$ through a quantum phase transition or crossover near $\mu_0$H$_C$$\sim$0.5 T into a polarized phase that is characterized by resolution-limited spin waves~\cite{Ross2009}.  Linear spin wave theory using anisotropic exchange produces an excellent description of the high field spin wave dispersion and intensities, and the resulting microscopic Hamiltonian has been used in high temperature series expansions which accurately describe the magnetization and heat capacity of Yb$_2$Ti$_2$O$_7$ in absolute units~\cite{Hayre2013}. Yb$_2$Ti$_2$O$_7$'s spin Hamiltonian contains 4 symmetry-allowed near-neighbour anisotropic exchange terms and it has been proposed that the largest of these is J$_{zz}$, which ferromagnetically couples together local z, or Ising components of spin~\cite{RossPRX}.  These z-components are aligned directly into or out of the tetrahedra, and this combination of S$_{\rm{eff}}=1/2$ spins, the spin Hamiltonian, and the pyrochlore lattice would then be responsible for the quantum spin ice phenomenology.  Somewhat different phenomenology has also been discussed in which J$_{zz}$ is less dominant~\cite{Robert2015}.

While the general phase behaviour for an anisotropic exchange Hamiltonian of the form which describes Yb$_2$Ti$_2$O$_7$ in zero magnetic field has exotic quantum spin liquid and Coulomb ferromagnetic mean field ground states present within it~\cite{SavaryPRL,SavaryPRB}, the mean field ground state predicted on the basis of Yb$_2$Ti$_2$O$_7$'s spin Hamiltonian is a simple ferromagnet with a mean field phase transition of T$_{\rm MF}$ $\sim$ 3 K~\cite{RossPRX}.  A broad ``hump" in Yb$_2$Ti$_2$O$_7$'s experimentally-determined heat capacity (C$_P$) is  observed at $\sim$ 2 K, roughly co-incident with the calculated T$_{\rm MF}$, however there are no indications of order observed above a sharp C$_P$ anomaly which occurs near T$_C$$\sim$ 0.26 K in stoichiometric powders of Yb$_2$Ti$_2$O$_7$~\cite{Hodges2002,Dalmas2006}.   Taking the sharp C$_P$ anomaly at face value for an indication of the ferromagnetic phase transition predicted by mean field theory, this indicates that T$_{\rm MF}$ is suppressed by a factor of $\sim$ 12 by quantum fluctuations, geometrical frustration, or both.  However, there are strong indications that Yb$_2$Ti$_2$O$_7$'s zero field phase below T$_C$ $\sim$ 0.26 K is far removed from a conventional ferromagnet.  To date there are experimental studies which support a relatively simple ferromagnetic ground state~\cite{Yasui2003,Chang2012}, but also studies which show extensive diffuse neutron scattering covering all of reciprocal space~\cite{Gardner2004}, no changes in the spin relaxation observed when cooling below the sharp C$_P$ anomaly in $\mu$SR studies from stoichiometric samples~\cite{Dortenzio2013}, and no evidence for the conventional spin waves that are expected as the normal modes of the magnetically-ordered state in any sample~\cite{Ross2009,Ross2011,Robert2015}.

An interesting feature of Yb$_2$Ti$_2$O$_7$'s exotic zero magnetic field ground state is its sensitivity to weak disorder~\cite{RossStuffing,Yaouanc2011}.  The sharp C$_P$ anomaly observed at T$_C$$\sim$0.26 K in stoichiometric powder samples, is observed to be broader and to occur at lower temperatures in all single crystal samples measured to date~\cite{Yaouanc2011}.  The C$_P$ anomalies can occur as low as $\sim$ 0.15 K in single crystals grown from floating zone image furnace techniques, may not obviously occur at any temperature, or may appear as multiple peaks at lower temperatures. 

In the interest of understanding the microscopic structure and defects responsible for the differences between powder samples grown by solid state synthesis, and single crystals grown from the melt by floating zone techniques, neutron diffraction studies were carried on both powder and crushed single crystal (CSC) samples, that were known to display different C$_P$ behaviour at low temperature~\cite{RossStuffing}.  The conclusions were that, while the powder sample was stoichiometric Yb$_2$Ti$_2$O$_7$, the CSC was characterized as exhibiting weak ``stuffing" wherein a small proportion of excess Yb$^{3+}$ resides on the Ti$^{4+}$ site, and the composition of the crushed single crystal was Yb$_{2+x}$Ti$_{2-x}$O$_{7+y}$ with $x=0.046$.  Note that this defect level is close to the limit of detectability by conventional diffraction techniques.  This study was greatly aided by the fact that Ti displays a negative coherent neutron scattering length, hence there is significant neutron contrast for Yb, which has a positive coherent neutron scattering length,  occupying the Ti site.  In and of itself, sensitivity of the ground state to such weak disorder is a remarkable result, as conventional three dimensional ordered states are not sensitive to disorder at such a low level.  For example, the phase transition to non-collinear $\Psi_2$ antiferromagnetic order in Er$_2$Ti$_2$O$_7$ is not sample dependent, and has been studied as a function of magnetic dilution and shown to be consistent with conventional three dimensional percolation theory~\cite{Niven2014}.  A recent study of the Yb$^{3+}$ crystal field excitations in Yb$_2$Ti$_2$O$_7$ suggests that the anisotropy of Yb$^{3+}$ in defective environments is Ising-like, rather than XY-like in stoichiometric Yb$_2$Ti$_2$O$_7$, and this may be related to the effectiveness of disorder at this low $\sim$ 1 $\%$ level~\cite{Gaudet2015}.

It is therefore important to fully characterize powder samples of Yb$_2$Ti$_2$O$_7$ which are known to be stoichiometric and to display sharp C$_P$ anomalies at T$_C$=0.26 K; that is to use this as a benchmark for understanding the zero field ground state of pristine Yb$_2$Ti$_2$O$_7$.  In this work, we report elastic and inelastic neutron scattering results from the same two powder samples previously studied by neutron diffraction~\cite{RossStuffing} and which are known to be stoichiometric Yb$_2$Ti$_2$O$_7$ and the CSC with composition Yb$_{2+x}$Ti$_{2-x}$O$_{7-y}$ with x=0.046.  In addition, we report a comparison between the spin dynamics measured on a stoichiometric powder of Yb$_2$Ti$_2$O$_7$, the CSC powder, a single crystal of Yb$_2$Ti$_2$O$_7$ grown by the floating zone technique, and a powder sample of Yb$_2$Ti$_2$O$_7$'s sister pyrochlore magnet, Yb$_2$Sn$_2$O$_7$.  We show that stoichiometric Yb$_2$Ti$_2$O$_7$ displays resolution-limited magnetic Bragg scattering at low temperatures, however these persist to a much higher temperature scale than T$_C$=0.26 K.  Nonetheless, looking only at the low temperature Bragg intensities, we refine a static magnetic structure which is best described as an ice-like splayed ferromagnet with an ordered moment of 0.90(9)$\mu_B$.  The spin dynamics at the lowest temperatures in stoichiometric Yb$_2$Ti$_2$O$_7$ in zero field are indeed far removed from conventional spin waves.  They are gapless on an energy scale $<$ 0.09 meV at all wave vectors, and characterized by a continuum of scattering with a bandwidth of $\sim$ 1 meV.  Vestiges of very overdamped ferromagnetic spin waves can be seen in the inelastic scattering, which itself is temperature independent below T$_{\rm MF}$$\sim$ 3 K.

\section{EXPERIMENTAL DETAILS}

The powder samples employed in this study were the same samples previously studied by Ross et al~\cite{RossStuffing}.  10 grams of stoichiometric Yb$_2$Ti$_2$O$_7$ and 8 grams of the CSC powder, Yb$_{2+x}$Ti$_{2-x}$O$_{7+y}$ with $x=0.046$, were separately loaded in aluminum sample cans with copper lids under 10 atms of helium exchange gas.  This method of loading powder samples is known to provide good thermal contact to the cold finger of the dilution refrigerator, and maintains thermal conductivity below 1K by enabling enough superfluid Helium to coat the powder grains~\cite{Ryan2009}. The 4 gram single crystal of Yb$_2$Ti$_2$O$_7$ (produced in a floating zone image furnace following similar procedures to those used to grow other single crystal titanate pyrochlores) was mounted on an aluminum holder.

Elastic neutron scattering measurements were carried out on the SPINS triple axis instrument at the NIST Center for Neutron Research.  SPINS used pyrolytic graphite as both monochromator and analyzer, and 80' collimators before and after the sample, producing a 5.0 meV elastically scattered beam with an energy resolution of 0.25 meV.  Cooled beryllium filters placed before and after the sample helped to eliminate higher harmonic contamination. The sample was first cooled to $T$=8K where scans of the (111), (002), (220), (113) and (222) Bragg peaks were collected. Afterwards, the sample was cooled down to 100mK and sat at this temperature for 2 hours before collecting measurements at the same Bragg positions as above. Finally, the sample was warmed up to 700mK where another scan of the same Bragg peaks was also collected.

Inelastic neutron scattering measurements were also performed on both the stoichiometric powder as well as the CSC, stuffed powder using the time-of-flight disk chopper spectrometer (DCS) at NIST~\cite{DCS}. For these measurements, monochromatic incident neutrons of wavelength of 5$\angstrom$ were employed, giving an energy resolution at the elastic position of 0.09 meV. Empty can measurement have been used as a background for these data. Inelastic neutron scattering measurements on the single crystal of Yb$_2$Ti$_2$O$_7$ were performed using the cold neutron chopper spectrometer (CNCS)~\cite{CNCS} at the Spallation Neutron Source (SNS) at Oak Ridge National Laboratory. For this time-of-flight experiment, the single crystal was mounted in a dilution refrigerator and aligned with the [HHL] plane in the horizontal plane. An incident neutron energy of 3.3 meV was employed which gave an energy resolution of ~ 0.1 meV at the elastic position. Sample rotation methods were employed wherein the single crystal sample was rotated 360$\degree$ about the vertical direction in 1$\degree$ steps. The background is approximated using a measurement done under the same conditions but without the sample. Magnetic and structural refinement were performed using SARAh Refine~\cite{SARAh} and FullProf~\cite{FullProf}.

\section{Results and Discussion}

\subsection{Elastic neutron scattering and magnetic structure determination:}

\begin{figure}[h]
\includegraphics[width=8.5cm]{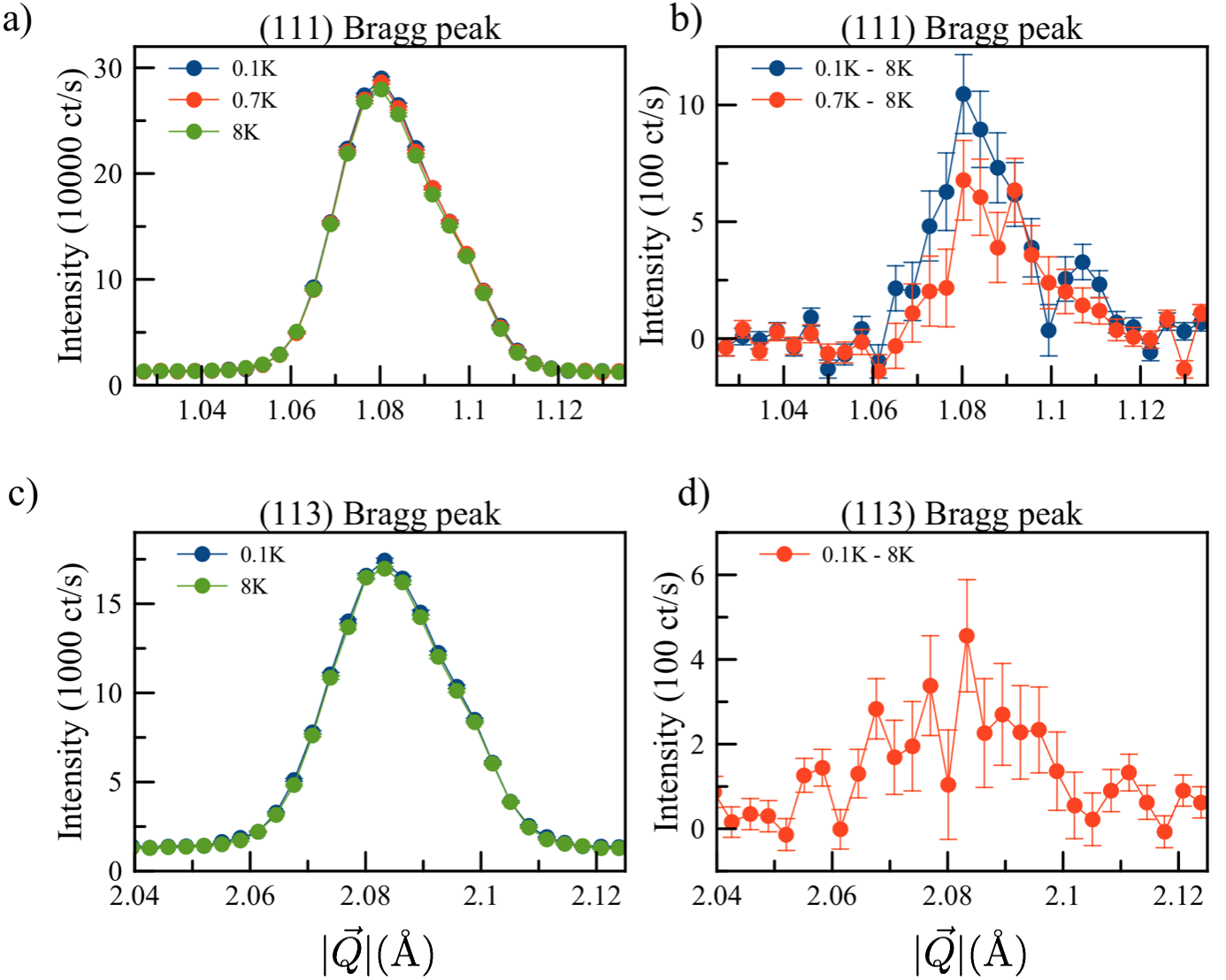}
\label{Bragg} 
\caption{\emph{ The temperature dependence of the (111) and (113) Bragg peaks.--} Panels a) and c)  show the elastically scattered neutron intensity as a function of the momentum transfer $|\vec{Q}|$ around the (111) and (113) Bragg peaks respectively. Panels b) and d) shows the subtraction between the different $|\vec{Q}|$ cuts at different temperatures. A clear, $|\vec{Q}|$-resolution-limited increase of the elastic scattering at these two Bragg peaks is observed and is representative of the data used to model the magnetic structure of stoichiometric Yb$_2$Ti$_2$O$_7$.  Error bars represent one standard deviation.}
\end{figure}

Fig.1 shows elastic neutron scattering data taken on the stoichiometric powder sample of Yb$_2$Ti$_2$O$_7$ using SPINS.  No magnetic Bragg peaks were observed at nuclear-disallowed positions upon cooling below $T_C$, and consequently we focused our attention on nuclear-allowed Bragg peaks typical of Q=0 magnetic structures. Fig.1 a) and b) show the (111) Bragg peak at three temperatures: T = 0.1 K (below $T_C$), T=0.7 K (above $T_C$), and T=8 K, which is above both the T$_{\rm MF}$ calculated for Yb$_2$Ti$_2$O$_7$ on the basis of its microscopic spin Hamiltonian, as well above the higher temperature ``hump" in its C$_P$ near $\sim$ 2 K.  Fig.1 c) shows the corresponding elastic scattering at the (113) Bragg position at T=0.1 K and T=8 K. Elastic scattering at both wave vectors is dominated by nuclear contributions, but a weak temperature dependent magnetic contribution is identified as can be seen in Fig.1 b) and d) which show  T=0.1 K $-$ T=0.7 K and T=0.7 K $-$ T=8 K data sets for (111) and a T=0.1 $-$ 8 K data set for (113), respectively.  Similar elastic scattering measurements were also carried out at the (220) Bragg position. The magnetic intensities extracted from these differences are listed in Table I, where we see that for all of the (111), (113) and (222) Bragg positions, the increase in scattering, relative to T=8 K, is 3.2(4) $\%$, 3.3(5) $\%$, and 3.6(6) $\%$, respectively.  Relative to T=0.7 K, the magnetic intensity at T=0.1 K is increased by only $\sim$ 1.3 $\%$ at the (111) Bragg position. We also note that the differences in the elastic scattering have the same Q-lineshape as the original Bragg scattering, implying that the corresponding magnetic order is long range, with a correlation length exceeding 80 $\angstrom$.

\begin{figure}[h]
\includegraphics[width=8.5cm]{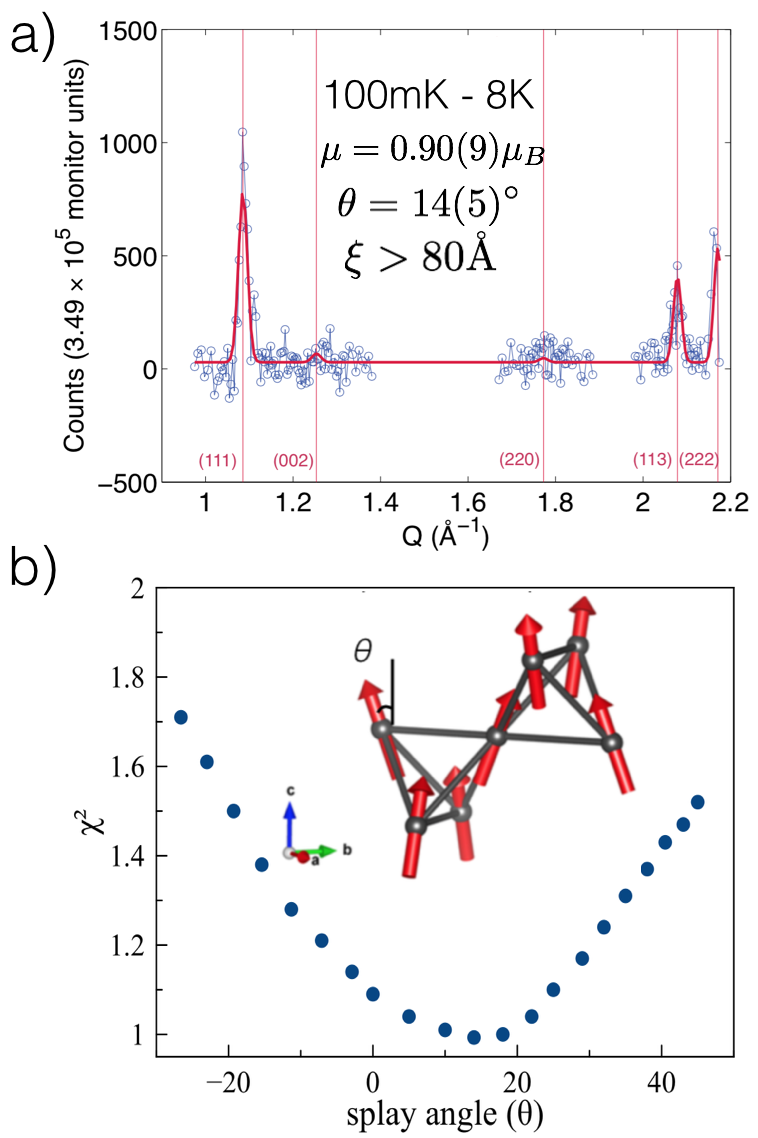}
\label{FullDiffpattern} 
\caption{\emph{Intensity of elastic magnetic scattering vs. $Q$ and the corresponding fit to the splayed Ice ferromagnet --} a) the best fit of the model for a splayed Ice ferromagnet to the difference of elastic neutron scattering intensity at 100 mK and 8 K (i.e., the magnetic elastic scattering) is shown.  The model employed is a splayed ferromagnet (a $Q$=0 structure with a net moment along the cubic axes) on the pyrochlore lattice, with splay angle of $\theta= 14(5)^{\circ}$ and an ordered moment of 0.90(9)$\mu_B$. Note: the error bars are not shown in this figure for clarity, but are taken into account as weights in the least squared fitting of the model. b) the goodness-of-fit parameter, $\chi^2$, as a function of splay angle is shown.  The inset to b) shows an illustration of the best fit magnetic structure model on a pair of tetrahedra.} 

\end{figure}

Ignoring for the time being the fact that the magnetic order parameter does not go to zero above T$_C$=0.26 K, we can nonetheless refine a magnetic structure based on the relative intensities of the differences in Bragg scattering between T=0.1 K and 8 K.  This difference is plotted as a function of $|Q|$ in Fig.2 a), along with the best fit to a model for a canted or splayed ferromagnet of the form depicted in the inset of Fig.2 b).  This generalized structure has the Yb moments along a [100] direction, but also allows for a canting angle, $\theta$, which can be either towards (+) or away from (-) the local [111] direction, that is the direction pointing into or out of the tetrahedron.  A collinear [100] ferromagnet would correspond to a canting angle of zero, while the ordered spin ice ground state, in which all moments point directly into or out of the tetrahedra, would correspond to $\theta$=54.7$^{\circ}$.  A negative canting angle gives an XY splayed ferromagnet. The best fit, shown as the solid line in Fig.2 a), corresponds to a splayed ice-like ferromagnetic ground state with a positive canting angle of 14 $\pm$ 5$^{\circ}$, and an ordered moment (relative to 8 K) of 0.90(9) $\mu_B$. The saturated moment determined by DC magnetization is 1.75 $\mu_B$ which compared to the order moment we obtained gives a spin polarization of 51 $\%$. Note that, as shown in Fig.1 b), most of the decrease in the Bragg scattering on raising the temperature occurs above 0.7 K, more than a factor of two above T$_C$=0.26 K. Our refinements of these magnetic structures are summarized in Table I, where we have calculated the expected increases in Bragg scattering at each of (111), (113) and (222) for two different Q=0 magnetic structures. The splayed XY ferromagnet, with a negative canting angle, is known to describe Yb$_2$Ti$_2$O$_7$'s sister pyrochlore magnet Yb$_2$Sn$_2$O$_7$ ($\theta$ $\sim$ -10$\degree$)~\cite{Yaouanc2013}, while the splayed spin ice structure has been reported in some studies of single crystal Yb$_2$Ti$_2$O$_7$ but with a smaller splay angle ($\theta$ $\sim$ 1$\degree$)~\cite{Chang2012,Yasui2003}. Fig.2 b) shows how the goodness of fit parameter ($\chi^2$) vary as a function of splay angle for this model of splayed or canted ferromagnets.  

Fig.3 shows the temperature dependence of the (111) Bragg intensity from the stoichiometric powder sample of Yb$_2$Ti$_2$O$_7$, relative to T=0.7 K.  This data set was collected while warming from 100mK to 700mK with a warming rate of 1 mK/min. For reference, we overploted the elastic scattering perviously measured near the (111) position in single crystal Yb$_2$Ti$_2$O$_7$ (note: both curves are scaled to their intensities at 100 mK, after subtracting their intensities at 700 mK). It is clear that these temperature dependencies mirror each other.  We also included C$_P$ as measured on a powder sample of Yb$_2$Ti$_2$O$_7$ with T$_C$=0.26 K.  One can see that these magnetic intensities do not resemble a conventional order parameter with T$_C$=0.26 K. The elastic intensities are approximately constant up to $\sim$ 0.35 K, and then begin to decrease.

\begin{figure}[h]
\includegraphics[width=8.5cm]{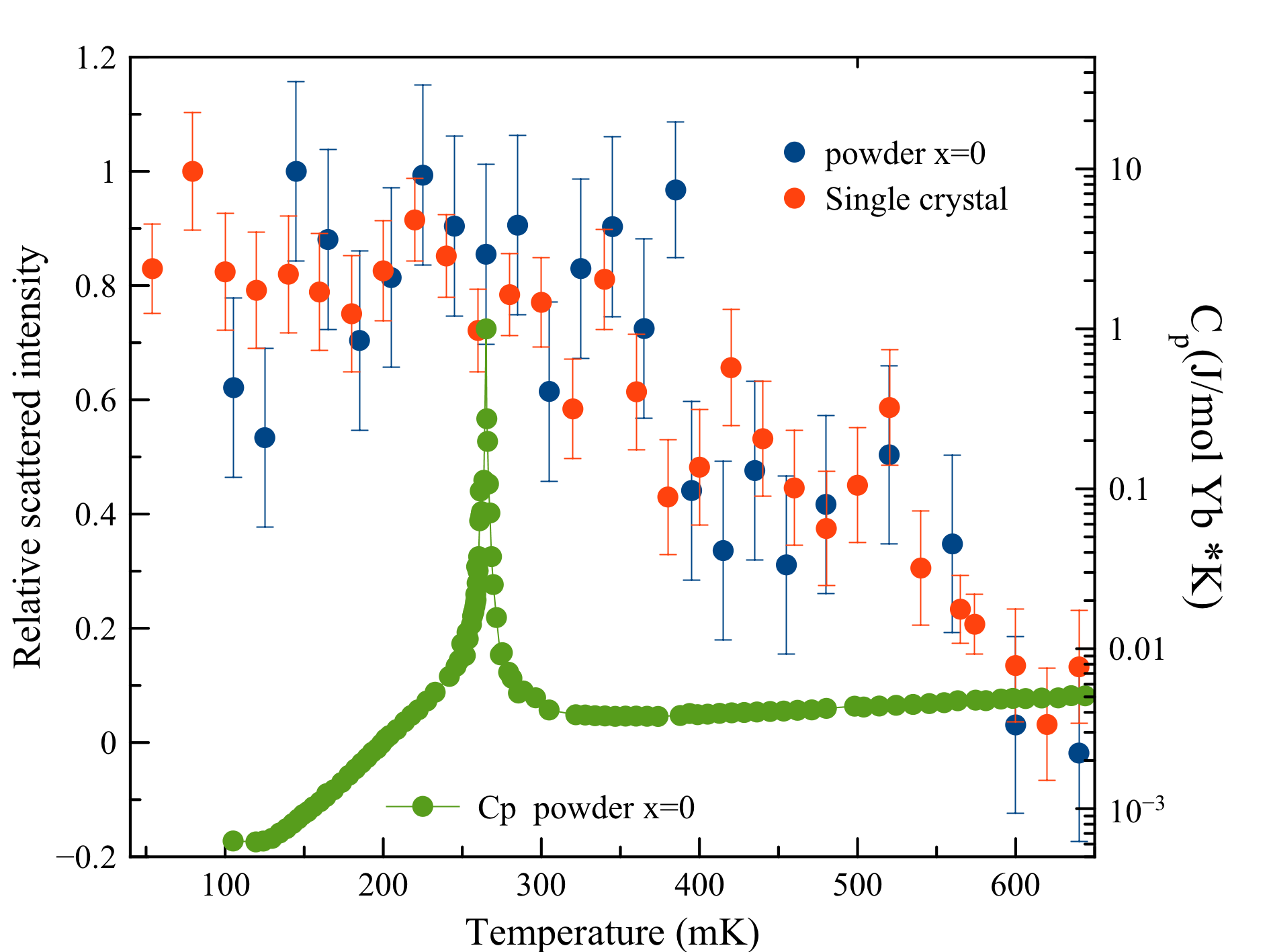}
\label{Tdep111} 
\caption{\emph{The temperature dependence of the elastic scattered intensity at $|\vec{Q}|$=(111).--} The scattered elastic neutron intensity, relative to T=700mK at $|\vec{Q}|$=(111) shows similar temperature dependence for both a previously measured single crystal (data taken from Ross et al.~\cite{Ross2009}) and for the stoichiometric powder sample of Yb$_2$Ti$_2$O$_7$ (this work). With increasing temperature, the elastic magnetic scattering begins to fall around $T$ = 0.35 K, and decreases approximately linearly above this temperature. The specific heat anomaly T$_C$$\sim$ 0.26 K for a representative powder sample of Yb$_2$Ti$_2$O$_7$ (data taken from ref Ross et al.~\cite{RossStuffing}) is shown for reference, and reveals the relative insensitivity of the temperature dependence of the elastic scattering to the specific heat anomaly.}
\end{figure}

\begin{table*} %add [H] placement to break table across pages
\begin{tabular}{|c||c|c|c|c|c|c|c|}
\hline\hline
 & $\%$ increase (111) & $\%$ increase (113) & $\%$ increase (222) & Moment Size & canting angle ($\theta$) & $   \chi^2$ \\
\hline
Measured & 3.2(4) & 3.3(5) & 3.6(6) & - & -  & - \\
Collinear FM & 3.2 & 4.5 & 5.6 & 1.1$\mu_B$ & 0$^o$ & 1.09\\
Splayed Ice FM & 3.2 & 3.1 & 4.2 & 0.90(9)$\mu_B$ & 14$^o$ $\pm$ 5$^o$ & 0.99\\
\hline\hline
\end{tabular}
\label{tab: ElasticInt}
\caption{\emph{A Comparison of the excess intensity in the measured Bragg peaks with model calculations for possible ferromagnetic ground states.--} The percent excess elastic intensity for the (111), the (113) and the (222) Bragg peaks are compared with those calculated for a (100) collinear ferromagnet and a Splayed Ice ferromagnet (Fig.2 b)). The best agreement between the observed excess intensity and such a model for canted ferromagnets is achieved with the Splayed Ice ferromagnet, and a splay angle of 14 $\pm$ 5$^\circ$.The magnetic moment size at the Yb$^{3+}$ site, based on the increase in Bragg intensity relative to 8 K, is also given for the proposed ordered states.
 }
\end{table*}

\subsection{Inelastic Neutron scattering and gapless spin excitations in the Yb$_2$Ti$_2$O$_7$ ground state:}

Inelastic neutron scattering measurements were carried out on our three Yb$_2$Ti$_2$O$_7$ samples (stoichiometric powder, lightly stuffed CSC powder, and single crystal) using time-of-flight neutron spectrometers.  Measurements were carried out on the stoichiometric Yb$_2$Ti$_2$O$_7$ powder sample and the CSC lightly stuffed powder with composition Yb$_{2+x}$Ti$_{2-x}$O$_{7+y}$  $x=0.046$ using the DCS direct geometry chopper spectrometer at NIST.  The resulting $S(|\vec{Q}|,E)$ for these two samples at T=0.1 K are shown in Fig.4 a) and b). The corresponding [HHL] plane-averaged data set (approximating a powder average) for single crystal Yb$_2$Ti$_2$O$_7$ at T=0.1 K is shown in Fig.4 c).   

\begin{figure}[h]
\includegraphics[width=8.5cm]{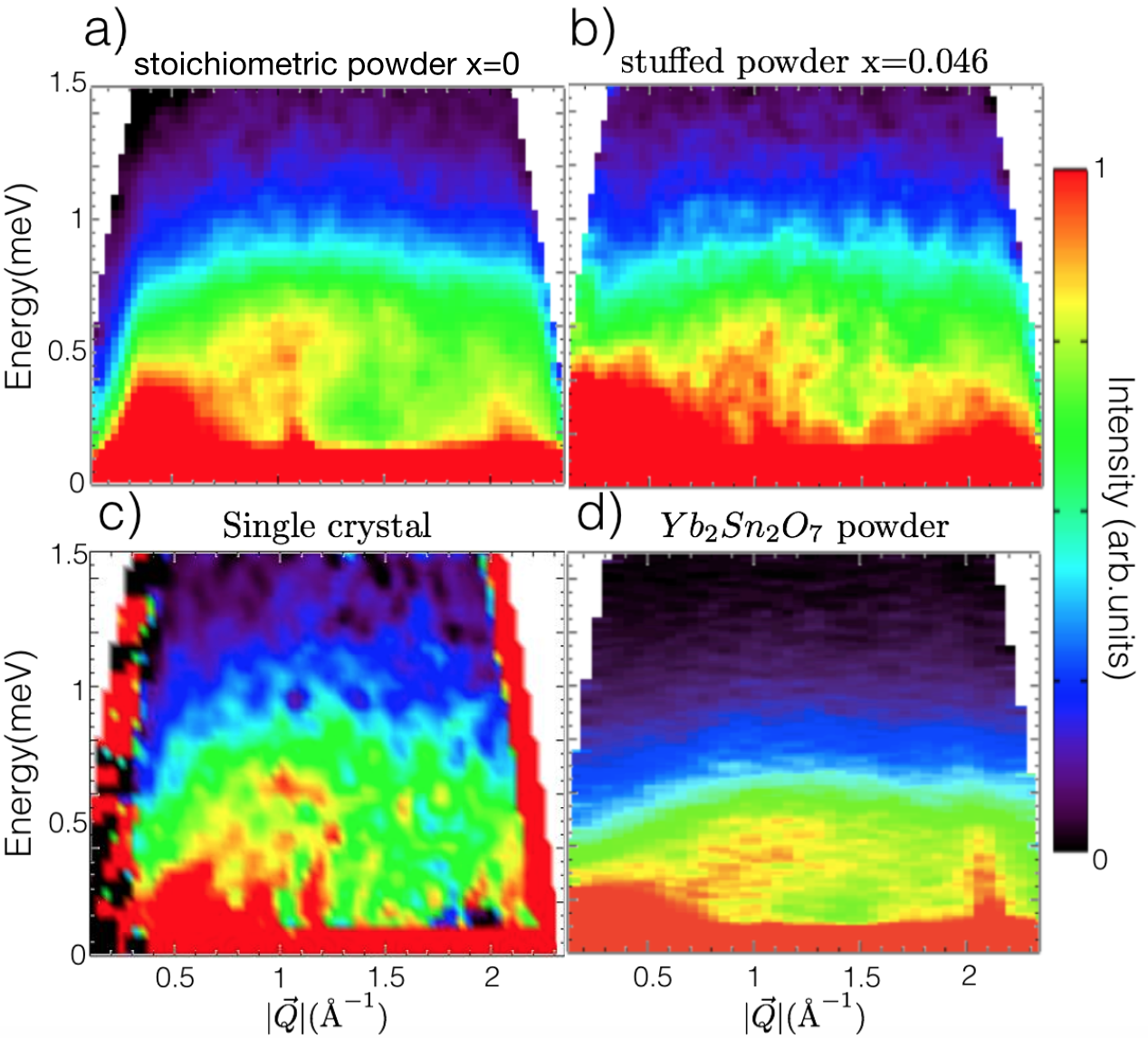}
\label{Eslice}
\caption{\emph{$S(|\vec{Q}|,E)$ as measured for stoichiometric and weakly stuffed Yb$_2$Ti$_2$O$_7$ with comparison to Yb$_2$Sn$_2$O$_7$.--} a) and b) show $S(|\vec{Q}|,E)$ measured on the stoichiometric Yb$_2$Ti$_2$O$_7$ powder and the lightly stuffed CSC powder of Yb$_{2+x}$Ti$_{2-x}$O$_{7+y}$ with $x$=0.046, respectively, both at T=0.1 K.  c) shows the corresponding $S(|\vec{Q}|,E)$ measurement on single crystal Yb$_2$Ti$_2$O$_7$, also at T=0.1 K.  Panel d) shows $S(|\vec{Q}|,E)$ measured on powder Yb$_2$Sn$_2$O$_7$ at T=0.1 K, by Dun \emph{et al}.\cite{Dun2013}. All four data sets have had an empty can background data set subtracted from them,  and have had intensities scaled such that the scattering at $\vec{|Q|}$ = 1$\angstrom^{-1}$ and E=0.5 meV saturates the colour scale.}
\end{figure}

\begin{figure}[h]
\includegraphics[width=8.5cm]{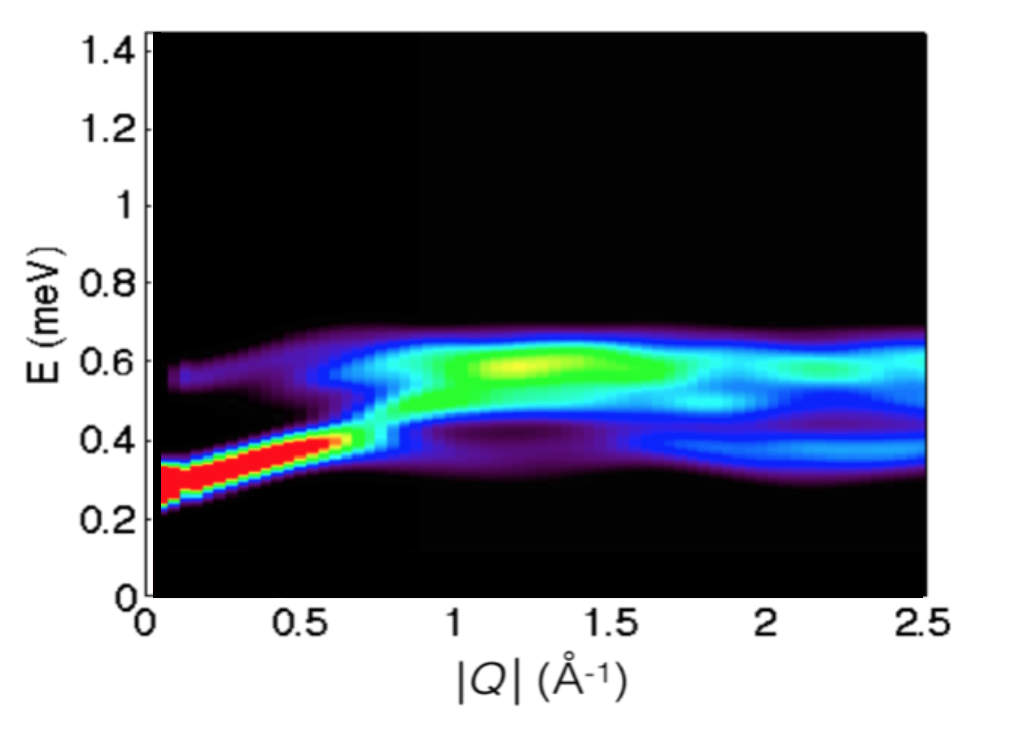}
\label{SpinWave}
\caption{\emph{Energy vs. $|\vec{Q}|$ slice of the mean-field calculation of $S(|\vec{Q}|,E)$ for Yb$_2$Ti$_2$O$_7$.--} $S(|\vec{Q}|,E)$ has been computed using an anisotropic spin 1/2 exchange Hamiltonian with the exchange parameters determined by Ross et al.\cite{RossPRX}. This calculation successfully accounts for the spin wave spectrum in the high magnetic field, polarized state at all wavevectors, but does not resemble the inelastic neutron scattering seen in Fig. 4.  In particular the measured inelastic magnetic scattering is gapless at all wavevectors, while the calculated $S(|\vec{Q}|,E)$, shown here, possesses a minimum gap of $\sim$ 0.25 meV.}
\end{figure}

\begin{figure*}[hbtp]
\includegraphics[width=16cm]{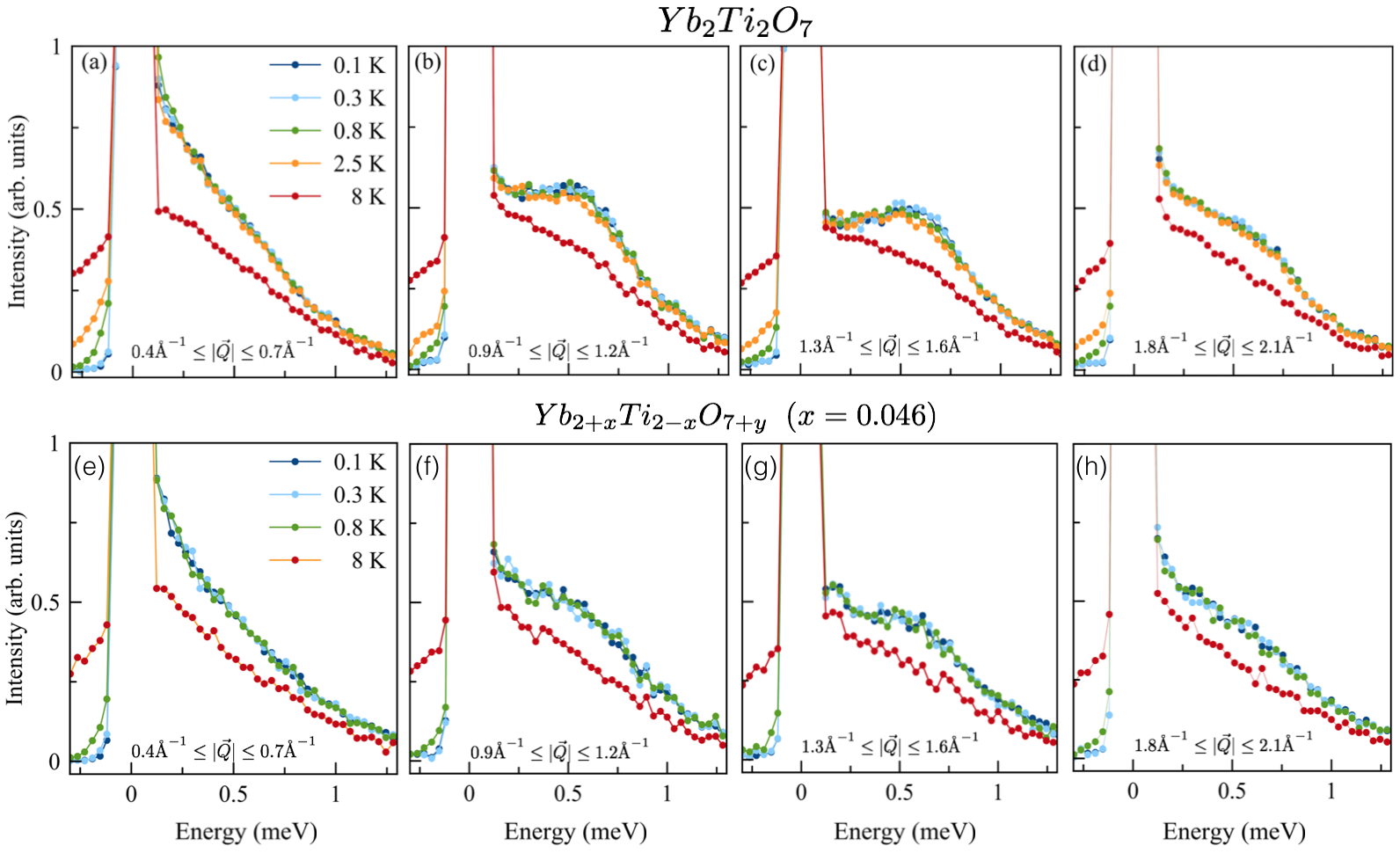}
\label{EnergyCuts}
\caption{\emph{Energy cuts of $S(|\vec{Q}|, E)$ for the stoichiometric powder sample of Yb$_2$Ti$_2$O$_7$ and the CSC Yb$_{2+x}$Ti$_{2-x}$O$_{7+y}$ with x=0.046.--} Top panels : These 4 different panels show cuts through $S(|\vec{Q}|, E)$ of Yb$_2$Ti$_2$O$_7$, taken from Fig. 4 a), for different ranges of $|\vec{Q}|$ and at different temperatures. Four different integrations in $|\vec{Q}|$ going from smaller $|\vec{Q}|$ in a) to larger $|\vec{Q}|$ in d) have been applied, as indicated in the individual panels. No differences in the scattering is observed from T=2.5 K down to T=0.1 K. Bottom panels : Same data and |$\vec{Q}$| integrated cuts as in a) - d) but for Yb$_{2+x}$Ti$_{2-x}$O$_{7+y}$ with x=0.046 taken from Fig. 4 b).}
\end{figure*}

The inelastic spectral weight shown in Fig.4 is known to be magnetic in origin from its previously determined field dependence, and from its temperature dependence to be discussed shortly.  We see that $S(|\vec{Q}|,E)$ in stoichiometric Yb$_2$Ti$_2$O$_7$ at T=0.1 K is largely characterized by a continuum of scattering below an upper band edge of $\sim$ 1 meV, and it is gapless on a energy scale $<$0.09 meV at all wave vectors measured.  The inelastic scattering has some weak structure to it, resembling overdamped ferromagnetic spin waves which disperse as $\hbar\omega \sim Q^2$ and whose intensity peaks at small $Q$. This continuum spin excitation spectrum is very different from that expected for spin waves within the ferromagnetically ordered state predicted by mean field theory on the basis of the anisotropic exchange Hamiltonian determined previously from the high field, polarized state. Fig.5 shows the powder average of this theoretical spin wave spectrum broadened by the instrumental resolution, and one clearly expects sharp and dispersing excitations. The predicted spin wave spectrum is gapped at all wave vectors, with a minimum gap of $\sim$ 0.25 meV at $Q$=0. Such a gap would be easily observed with our inelastic measurements using DCS. 

While the spin excitation spectrum in stoichiometric Yb$_2$Ti$_2$O$_7$ is exotic, and quite distinct from the expectations of anisotropic spin wave theory, comparison to the measured spectrum in lightly stuffed CSC Yb$_{2+x}$Ti$_{2-x}$O$_{7+y}$ with $x$=0.046, in Fig.4 b), and to single crystal Yb$_2$Ti$_2$O$_7$, in Fig.4 c)  which is also likely lightly stuffed, shows that is not particularly sensitive to weak disorder.  The upper band edge of the continuum slightly softens in the lightly stuffed CSC sample and single crystal, and the vestiges of the overdamped ferromagnetic spin waves are not as clear compared with the stoichiometric Yb$_2$Ti$_2$O$_7$ in Fig. 4 a).  However, the continuum nature of the magnetic scattering and its lack of a gap at any wave vector are common to all three samples.  

These features are also common to $S(|\vec{Q}|,E)$ measured on DCS under $\sim$ identical conditions on a powder sample of Yb$_2$Sn$_2$O$_7$ by Dun and co-workers shown in Fig.4 d).  These measurements on Yb$_2$Sn$_2$O$_7$ were previously published~\cite{Dun2013}, albeit using a high temperature subtraction, as opposed to the empty can subtraction we employ here.  Yb$_2$Sn$_2$O$_7$ cannot be grown as a single crystal due to the volatility of the SnO$_2$ starting materials, which also likely implies some level of defects in its powders.   

Energy cuts through the $S(|\vec{Q}|,E)$ data sets for stoichiometric powder Yb$_2$Ti$_2$O$_7$ and lightly stuffed CSC with approximate composition Yb$_{2+x}$Ti$_{2-x}$O$_{7+y}$ and $x$=0.046 are shown in Figs.6.  While differences between the four data sets of Fig.4 can be seen at $Q$ $\leq$ 0.4 $\angstrom^{-1}$, this low-$Q$ region corresponds to the edge of the neutron detector bank closest to the incident beam, and is thus the part of the data set that is most sensitive to the precise details of the background subtraction. This low-$Q$ region of scattering is thus avoided in the following quantitative analysis shown in the energy scans of Fig.6. These energy scans employ different $Q$-integrations, going from smaller $Q$ to larger $Q$ in panels Fig.6 a) - d) for Yb$_2$Ti$_2$O$_7$ and Fig.6 e) - h) for Yb$_{2+x}$Ti$_{2-x}$O$_{7+y}$ and $x$=0.046.  Consistent with the detailed $S(|\vec{Q}|,E)$ maps, we see that pronounced inelastic shoulders are observed at intermediate $Q$'s, as seen in Figs.6 b) and c), and these are the vestiges of the ferromagnetic spin waves discussed earlier.  Well defined ferromagnetic spin waves would show inelastic peaks which would disperse as $E \sim Q^2$.  As seen on Fig.6 f) - g), the shoulders are considerably more rounded in the lightly stuffed CSC sample, indicating that the vestiges of these overdamped excitations are even further overdamped in the presence of weak disorder.

The energy cuts through $S(|\vec{Q}|,E)$ in Fig.6, also shows the temperature evolution of this inelastic scattering from T=0.1 K, well below T$_C$=0.26 K, to T=8 K, above T$_{\rm MF}$ and the ``hump" in C$_P$ near 2 K.  This temperature dependence is very similar for both the stoichiometric Yb$_2$Ti$_2$O$_7$ and the lightly stuffed CSC powder sample, so we focus our discussion on the stoichiometric Yb$_2$Ti$_2$O$_7$ energy cuts of $S(|\vec{Q}|,E)$ shown on the top panels of Fig.6.  It is worth noting that the temperature evolution of $S(|\vec{Q}|, E)$ on the negative energy side, shows clear evolution of the scattering from T=0.8 to T=0.3 K, implying that the stoichiometric Yb$_2$Ti$_2$O$_7$ was equilibrated down to at least 0.3 K.  While not easily visible in Fig.6, there is a continued decrease in the inelastic scattering near -0.1 meV from T=0.3 K to 0.1 K, strongly suggesting that the system equilibrated at all temperatures measured. 

The most striking feature of the temperature dependence of $S(|\vec{Q}|,E)$ in Yb$_2$Ti$_2$O$_7$ is that there is none for temperatures less than $\sim$ 2.5 K. That is, the gapless, continuum form of $S(|\vec{Q}|,E)$ in stoichiometric Yb$_2$Ti$_2$O$_7$ is maintained through its large C$_P$ anomaly at T$_C$=0.26 K, up to 10 times this temperature.  Changes in $S(|\vec{Q}|,E)$ only appear in the temperature range from 2.5 K to 8 K, consistent with a temperature evolution on the scale of the calculated T$_{\rm MF}$ $\sim$ 3 K, or the high temperature "hump" observed in T$_C$ near 2 K.  Very similar temperature dependence of $S(|\vec{Q}|,E)$ is observed for the lightly stuffed CSC sample, as shown in the bottom panels of Fig.6.

\section{Conclusions}

New neutron scattering measurements on a stoichiometric powder sample of Yb$_2$Ti$_2$O$_7$ reveal low temperature magnetic Bragg peaks, overlapping with nuclear-allowed Bragg peaks. This elastic scattering is interpreted in terms of a long range ordered, splayed spin ice static structure with a correlation length exceeding 80$\angstrom$  and an ordered moment, relative to 8 K, of 0.90(9) $\mu_B$. However, the temperature dependence of the elastic Bragg scattering does not correlate with the T$_C$=0.26 K expected for such powder samples. Rather, this elastic scattering begins to dissipate above $\sim$ 350 mK and shows continued decrease above 700 mK.

The zero field magnetic inelastic spectrum of stoichiometric Yb$_2$Ti$_2$O$_7$ is exotic. It shows a gapless, continuum form, at T=0.1 K, well below T$_C$=0.26 K, with an upper band edge of $\sim$ 1 meV.  No spin gap is observed at any wave vector to an upper limit of 0.09 meV, in contrast to the expectations of anisotropic spin wave theory within the ferromagnetically ordered state predicted from the mean field model using the spin Hamiltonian derived from high field spin wave measurements.  The continuum inelastic spectrum shows some weak structure, with vestiges of overdamped ferromagnetic spin waves present at small $Q$.  The inelastic magnetic spectrum in stoichiometric Yb$_2$Ti$_2$O$_7$ shows little or no temperature dependence up to temperatures greater than 2.5 K, much larger than T$_C$, and on the order of T$_{\rm MF}$ $\sim$ 3 K, and the temperature characterizing the high temperature ``hump" in C$_P$ near 2 K.  The form of the magnetic inelastic scattering, its continuum nature and temperature dependence, is only slightly influenced by weak stuffing, as is known to characterize the non-stoichiometric CSC powder sample and single crystal samples that were also measured.\\

\begin{acknowledgments}

We wish to thank Jan Kycia for collaborations and discussion related to heat capacity measurements. We thank C. Wiebe and H. Zhou for making their data available to this work. The neutron scattering data were reduced using Mantid~\cite{Mantid} and analyzed using DAVE software package~\cite{Dave}. Research using ORNL's Spallation Neutron Source was sponsored by the Scientific User Facilities Division, Office of Basic Energy Sciences, U.S. Department of Energy. The NIST Center for Neutron Research is supported in part by the National Science Foundation under Agreement No. DMR-094472. Works at McMaster University was supported by the National Sciences and Engineering Research Council of Canada (NSERC).  

\end{acknowledgments}

\bibliography{bibYTOPowd.bib}

\end{document}